\documentclass[prl,aps,twocolumn,showpacs]{revtex4}
\usepackage[dvipdfm]{graphicx}
\usepackage{dcolumn}
\usepackage{bm}
\usepackage{color}
\usepackage{ulem}
\usepackage{amssymb}
\usepackage{amssymb}
\usepackage{amsmath}

\begin{document}
\newcommand{\bR}{\mbox{\boldmath $R$}}
\newcommand{\tr}[1]{\textcolor{red}{#1}}
\newcommand{\trs}[1]{\textcolor{red}{\sout{#1}}}
\newcommand{\tb}[1]{\textcolor{blue}{#1}}
\newcommand{\tbs}[1]{\textcolor{blue}{\sout{#1}}}
\newcommand{\Ha}{\mathcal{H}}
\newcommand{\mh}{\mathsf{h}}
\newcommand{\mA}{\mathsf{A}}
\newcommand{\mB}{\mathsf{B}}
\newcommand{\mC}{\mathsf{C}}
\newcommand{\mS}{\mathsf{S}}
\newcommand{\mU}{\mathsf{U}}
\newcommand{\mX}{\mathsf{X}}
\newcommand{\sP}{\mathcal{P}}
\newcommand{\sL}{\mathcal{L}}
\newcommand{\sO}{\mathcal{O}}
\newcommand{\la}{\langle}
\newcommand{\ra}{\rangle}
\newcommand{\ga}{\alpha}
\newcommand{\gb}{\beta}
\newcommand{\gc}{\gamma}
\newcommand{\gs}{\sigma}
\newcommand{\vk}{{\bm{k}}}
\newcommand{\vq}{{\bm{q}}}
\newcommand{\vR}{{\bm{R}}}
\newcommand{\vQ}{{\bm{Q}}}
\newcommand{\vga}{{\bm{\alpha}}}
\newcommand{\vgc}{{\bm{\gamma}}}
\newcommand{\Ns}{N_{\text{s}}}
\newcommand{\vG}{{\bf G}}
\newcommand{\vx}{{\bf r}}
\newcommand{\vK}{{\bf K}}
\newcommand{\vp}{{\bf p}}
\newcommand{\vT}{{\bf T}}

\newcommand{\avrg}[1]{\left\langle #1 \right\rangle}
\newcommand{\eqsa}[1]{\begin{eqnarray} #1 \end{eqnarray}}
\newcommand{\eqwd}[1]{\begin{widetext}\begin{eqnarray} #1 \end{eqnarray}\end{widetext}}
\newcommand{\hatd}[2]{\hat{ #1 }^{\dagger}_{ #2 }}
\newcommand{\hatn}[2]{\hat{ #1 }^{\ }_{ #2 }}
\newcommand{\wdtd}[2]{\widetilde{ #1 }^{\dagger}_{ #2 }}
\newcommand{\wdtn}[2]{\widetilde{ #1 }^{\ }_{ #2 }}
\newcommand{\cond}[1]{\overline{ #1 }_{0}}
\newcommand{\conp}[2]{\overline{ #1 }_{0#2}}
\newcommand{\nn}{\nonumber\\}
\newcommand{\cdt}{$\cdot$}
\newcommand{\bra}[1]{\langle#1|}
\newcommand{\ket}[1]{|#1\rangle}
\newcommand{\braket}[2]{\langle #1 | #2 \rangle}
\newcommand{\bvec}[1]{\mbox{\boldmath$#1$}}
\newcommand{\blue}[1]{{#1}}
\newcommand{\bl}[1]{{#1}}
\newcommand{\bn}[1]{\textcolor{blue}{#1}}
\newcommand{\rr}[1]{{#1}}
\newcommand{\bu}[1]{\textcolor{blue}{#1}}
\newcommand{\red}[1]{{#1}}
\newcommand{\fj}[1]{{#1}}
\newcommand{\green}[1]{{#1}}
\newcommand{\gr}[1]{\textcolor{green}{#1}}
\definecolor{green}{rgb}{0,0.5,0.1}
\definecolor{blue}{rgb}{0,0,0.8}
\preprint{APS/123-QED}

\title{
{\it Ab initio} Studies on the Interplay between Spin-Orbit Interaction and Coulomb Correlation in Sr$_2$IrO$_4$ and Ba$_2$IrO$_4$ 
}
\author{R. Arita$^{1,2,3}$, J. Kune\v{s}$^{4}$, A. V. Kozhevnikov$^{5}$, A. G. Eguiluz$^{6}$, and M. Imada$^{1,3}$}
\affiliation{$^1$Department of Applied Physics, University of Tokyo, Hongo, Bunkyo-ku, Tokyo, 113-8656, Japan.}
\affiliation{$^2$JST-PRESTO, Kawaguchi, Saitama 332-0012, Japan.}
\affiliation{$^3$JST-CREST, Hongo, Bunkyo-ku, Tokyo, 113-8656, Japan.}
\affiliation{$^4$Institute of Physics, Academy of Sciences of the Czech Republic, Cukrovarnick\'a 10, 
Praha 6, 162 53, Czech Republic.}
\affiliation{$^5$Institute for Theoretical Physics, ETH Zurich, CH-8093 Zurich, Switzerland}
\affiliation{$^6$Department of Physics and Astronomy, The University of Tennessee, Knoxville, Tennessee 37996, USA}
\date{\today}

\begin{abstract}
{\it Ab initio} analyses of A$_2$IrO$_4$ (A=Sr, Ba) are presented.
Effective Hubbard-type models for Ir 5$d$ $t_{2g}$ manifolds downfolded from the global band structure 
are solved based on the dynamical mean-field theory. 
The results for A=Sr and Ba
correctly reproduce paramagnetic metals undergoing continuous transitions to insulators below the N\'eel temperature $T_N$. These compounds are classified not into Mott insulators but into Slater insulators. However, the insulating gap opens by a synergy of the N\'eel order and significant band renormalization, 
which is also manifested by a 2D bad metallic behavior in the paramagnetic phase near the
quantum criticality.
\end{abstract}

\pacs{71.30.+h,71.20.-b,71.15.-m}

\maketitle
\paragraph{-{\it Introduction.}}
Electron correlation effects in 
quasi-two-dimensional systems, 
especially transition metal compounds,
have been one of the central issues in condensed matter physics 
since the discovery of the cuprate superconductors. 
While the effective electron-electron interaction is large in the 3$d$ transition metal compounds, 
it 
becomes weaker in compounds built from heavier transition metal elements, e.g. 5$d$,
because of their more spatially extended orbitals.
On the other hand, the strength of the spin-orbit interaction (SOI) increases and thus 
in $5d$ compounds it is competitive with other characteristic energy scales:
The interplay between SOI, on-site repulsion, crystal-field splitting and inter-site hopping
opens a new field of research in strongly correlated compounds.

Ir compounds offer such model systems.
In particular, Sr$_2$IrO$_4$ and Ba$_2$IrO$_4$ are isostructural to
La$_2$CuO$_4$ implying a two-dimensional anisotropy\cite{Huang,Akimitsu}.
Since ten-fold degenerate Ir 5$d$ bands are partially occupied nominally by an odd number (five) electrons,
the insulating behavior of Sr$_2$IrO$_4$\cite{Moon}, with antiferromagnetic (AF) order below 250K\cite{Crawford,Vogt}, 
indicates an important role of electronic correlations. Based on the band structure of Sr$_2$IrO$_4$,
Kim {\it et al.}\cite{KimPRL} have shown that the $t_{2g}$ manifold near the Fermi level
is well separated from the $e_g$ one. 
SOI lifts the $t_{2g}$ degeneracy splitting it into four, mostly filled, degenerate states 
indexed with the angular momentum $j=3/2$, and a doubly degenerate 
$j=1/2$ state filled with nearly one electron.
They also suggested that low-energy excitations are 
accounted for by a single-band Hubbard model with two pseudospin 
states $j_z=\pm 1/2$ expressed by a linear combination of the three
$t_{2g}$ states;
$|j=1/2;j_z=\pm 1/2 \rangle =(|yz, \pm\sigma\rangle\mp i |zx,\pm \sigma \rangle \mp |xy,\mp\sigma\rangle)/\sqrt{3}$, 
where $\sigma$ describes the spin.
They proposed that Sr$_2$IrO$_4$ provides a realization of 
(weakly canted) AF Mott insulator similar to 
the parent materials of the high-$T_c$ cuprates.
Recent resonant X-ray scattering measurement also supports this scenario\cite{Takagi}. 

Ba$_2$IrO$_4$ 
is expected to have a narrower bandwidth because of the larger ionic radius
of Ba. However,
the AF transition around 240K is similar to the Sr compound\cite{Akimitsu}.

In this Letter, we present a detailed {\it ab initio} analysis of A$_2$IrO$_4$ (A=Sr, Ba). 
To predict accurately the electronic structure from the first principles, we employ a recently 
proposed three-stage 
scheme \cite{ImadaMiyake2010,Ferdi,Imai}.
First, we obtain the global band structure using the density functional theory (DFT) 
within the generalized gradient approximation (GGA)\cite{GGA}. 
Second, using a Wannier projection on the Ir $t_{2g}$ target bands
we eliminate the states far from the Fermi level. 
We employ the constrained random phase approximation (cRPA) \cite{Ferdi} to obtain the screened interaction 
parameters for the downfolded six-orbital Hubbard-type model. 
Third, we solve the derived low-energy ab initio model and study its dynamics with the dynamical mean-field theory (DMFT)\cite{DMFT}. This general {\it ab initio} framework may be called Multi-energy-scale {\it Ab initio} scheme for Correlated Electrons (MACE).MACE represents a general and hierarchical framework\cite{ImadaMiyake2010}, which enables {\it ab initio} calculations of strongly correlated electron systems. Among possible choices of MACE, here, we employ the combination of DFT and DMFT. 
 
Our results reveal that, in agreement with experiments,
Sr$_2$IrO$_4$ and Ba$_2$IrO$_4$ show a similar behavior, both with 
a continuous transition from a low temperature ($T$) AF insulator to a high-$T$ 
paramagnetic {\it metal}\cite{Kini}, 
but with a smaller energy scale (bandwidth, interaction strength) 
for Ba$_2$IrO$_4$.
Despite not being concomitant, the gap opening is induced by the AF transition,
in contrast to La$_2$CuO$_4$ but possibly similar to Nd$_2$CuO$_4$\cite{Kotliar}.
Therefore, we classify the iridates as Slater insulators.
The effective single-band and square-lattice behavior makes these materials valuable reference systems for the cuprate superconductors.
Indeed, we find their paramagnetic phases to be strongly renormalized 2D 
metals. 
Our results also provide insight into the proposed possible high $T_c$ superconductivity\cite{Wang-Senthil}
and giant magnetoelectric effects\cite{Chikara} in these compounds. 

\begin{figure}[htbp]
\begin{center}
\includegraphics[width=0.45\textwidth]{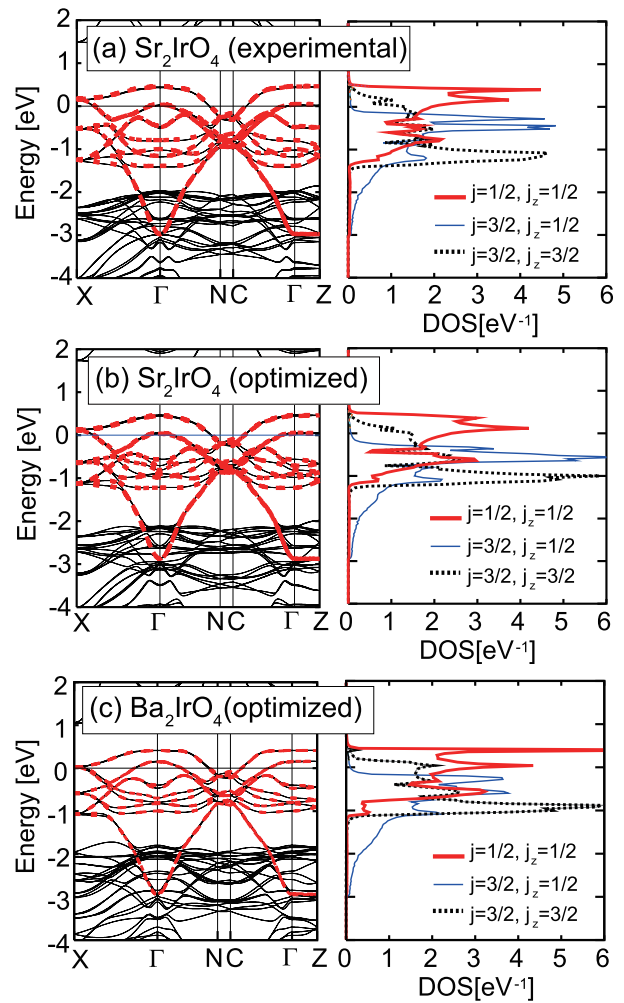}
\caption{
(Color online) (Left) GGA band structure (solid lines) together with the Wannier interpolated band of the $t_{2g}$ states ((red) dotted lines) for Sr$_2$IrO$_4$ (for the experimental structure (a) and optimized structure (b)) and Ba$_2$IrO$_4$ (c). (Right) Partial density of states for $j=1/2$ ((red) solid line), $j=3/2, j_z=\pm 1/2$((blue) thin line) and $j=3/2, j_z=\pm 3/2$((black) dotted line) states.
}
\label{Fig1}
\end{center}
\end{figure}

\paragraph{-{\it ab-initio downfolding.}}
In Fig.~\ref{Fig1}, we show the band structure of Sr$_2$IrO$_4$ and Ba$_2$IrO$_4$ obtained by the density functional calculation. 
We used the PBE exchange-correlation functional\cite{GGA} and the augmented plane wave and local orbital (APW+lo) method 
including the spin-orbit coupling as implemented in the WIEN2K program\cite{WIEN2k,wien_soc}. 
The muffin tin radii ($R_{\rm MT}$) of 2.23, 1.97, 1.74 bohr for Sr, Ir and O were used for Sr$_2$IrO$_4$ 
and 2.34, 1.92, and 1.70 bohr for Ba, Ir and O for Ba$_2$IrO$_4$, respectively. 
The maximum modulus for the reciprocal vectors $K_{\rm max}$ was chosen such that $R_{\rm MT}K_{\rm max}$ = 7.0 and
a 10 $\times$ 10 $\times$ 10 $k$-mesh in the first Brillouin zone was used. 

The crystal structure of Ref.\cite{Sr2IrO4-XtalStr} was used for Sr$_2$IrO$_4$. The obtained band structure is similar 
to previous studies\cite{KimPRL,LDA+U}. The experimental lattice parameters, $a$ and $c$, of Ba$_2$IrO$_4$ are 3\% and 4\% larger than those of Sr$_2$IrO$_4$\cite{Akimitsu}, while 
the Ir-O-Ir angle is yet to be determined experimentally. Therefore, we have optimized the structure requiring
any atomic force to be less than 0.2 mRy/bohr. 
The obtained Ir-O-Ir angle is 164$^\circ$. 
The difference between the optimized (153$^\circ$) and experimental (159$^\circ$) Ir-O-Ir angles for Sr$_2$IrO$_4$ suggests a few degrees overestimate of the Ir-O-Ir in Ba$_2$IrO$_4$ as well. 

We then constructed the Wannier functions for the $t_{2g}$-like bands, using the WIEN2Wannier \cite{W2W} and the wannier90 \cite{wannier90} codes. In Fig.~\ref{Fig1}, we superimpose the Wannier-interpolated bands on the original GGA bands. The hopping between the $d_{xy}$ orbitals is well represented by a two dimensional tight-binding model while the $d_{yz}$ and $d_{zx}$ bands are quite one-dimensional. The spin-orbit coupling mixes the cubic harmonics to form eigenstates of the pseudospin $j$. We used the $j$, $j_z$ basis, which diagonalizes the onsite part of the Hamiltonian and in which the local Green function is to a good approximation diagonal, in the subsequent calculations. The hopping amplitudes are listed in the Supplemental material. In the right panels of Fig.~\ref{Fig1}, we show the partial density of states (pDOS) of the $j = 1/2$ and $j = 3/2$ states. The center of gravity of the $j = 1/2$ bands is higher than that of $j = 3/2$ bands, and its contribution is dominant around the Fermi level. It is interesting to note that two one-dimensional bands ($d_{yz}$ and $d_{zx}$) and one two-dimensional band ($d_{xy}$) are mixed up equally and make quite two-dimensional $j=1/2$ bands. The width of the $j = 1/2$ band is about 1.5 eV, a few times smaller than that of the $d_{x^2-y^2}$ band of the cuprates. If we replace Sr with Ba, the unit cell expands, and the band width becomes smaller.

Next, we evaluate
the interaction parameters 
by cRPA\cite{Ferdi}. 
We used the Density Response Code (DRC)\cite{drc-code} recently developed for the Elk branch of the original Exciting FP-LAPW code\cite{exciting}. Below we sketch the main steps of the calculation. DRC utilizes 
the time dependent DFT to compute density susceptibility,
$
  \chi(\vx,\vx',\omega) =\chi^{KS}(\vx,\vx',\omega)+\iint d \vx_1 d \vx_2 \; \chi^{KS}(\vx,\vx_1,\omega) 
\big(v(\vx_1,\vx_2) + f^{xc}(\vx_1,\vx_2,\omega)\big) \chi(\vx_2,\vx',\omega).
$
Here $v(\vx_1,\vx_2)=|\vx_1-\vx_2|^{-1}$ is the bare Coulomb interaction and $f^{xc}$ is a dynamical exchange-correlation kernel ignored in
the conventional cRPA. 
\begin{equation} \label{eq:chi_ks} 
  \chi^{KS}(\vx,\vx',\omega)=\sum_{j,j'}' \frac{(f_j-f_{j'})\psi_{j'}(\vx) \psi_{j'}^{*}(\vx') \psi_j(\vx') \psi_j^{*}(\vx)} {\omega-(\epsilon_{j'} - \epsilon_j) + i0^{+}}, 
\end{equation}
is the susceptibility of the noninteracting Kohn-Sham electrons. Here, $\epsilon_{j}$ and $f_j$ are the energy and occupancy of the eigen-state $\psi_j$, and $\sum'$ runs over all pairs of bands but excludes the cases of $j,j'$ both belonging to the target $t_{2g}$ bands. 
Then the screened Coulomb interaction defined by
$
  W(\vx,\vx',\omega) = v(\vx,\vx') + 
  \iint d \vx_1 d \vx_2 \; v(\vx,\vx_1) \chi(\vx_1,\vx_2,\omega) v(\vx_2,\vx',\omega)
$
yields Hubbard-$U$ parameters (''4-index $U$'' matrix):
\begin{eqnarray} 
U_{\alpha \beta \gamma \delta} &=& \lim_{\omega \rightarrow 0}\iint d \vx_1 d \vx_2 \; w^{*}_{\alpha}(\vx_1)w^{*}_{\beta}(\vx_2)W(\vx_1,\vx_2,\omega) \times \nonumber \\
        &\times&   w_{\gamma}(\vx_1)w_{\delta}(\vx_2)
\label{eq:u_scr}
\end{eqnarray}
with Greek letters representing a combined index of band and translation. 
We used the band-disentanglement method\cite{disentanglecRPA} to unhybridize the $t_{2g}$ bands from the rest.

Here, to save the numerical cost, we constructed the simplified crystal structure with one formula unit per unit cell moving the inplane oxygen atoms to the symmetric positions. We took 100 unoccupied bands and $5 \times 5 \times 5$ $\vk$- and $\vq$-meshes in the calculation of the dielectric function. The double Fourier transform of $\chi(\vx,\vx',\omega)$ was done with the $|\vG+\vq|=3.5$ [1/a.u.] cutoff which gives $\sim$540 and $\sim$470 G-vectors for Ba$_2$IrO$_4$ and Sr$_2$IrO$_4$, respectively. 

In Table \ref{Table1}, we list the Coulomb and exchange interactions ($U_{ij} \equiv U_{ijij}$ and $J_{ij} \equiv U_{ijji}$, where $i$ and $j$ denote the $t_{2g}$ orbital index) in the cubic harmonics basis. From those we have deduced an approximate parametrization in terms of Slater parameter $F_0$, $F_2$, while fixing the ratio $F_4 / F_2$ to the atomic value of 0.625. We have found $F_0=1.933 (1.623)$ eV and $F_2=2.266 (2.068)$ eV well parametrize the interaction matrix for  Sr$_2$IrO$_4$ (Ba$_2$IrO$_4$). From these Slater parameters we have calculated the pairwise interactions $U_{ij}$  in the $j$, $j_z$ basis (see the Supplemental material). The expanded unit cell for Ba$_2$IrO$_4$ makes the crystal field splitting between the $t_{2g}$ and $e_g$ smaller and the screening becomes more efficient. This nearly cancels the band narrowing, leaving, the ratio of the band width to the interaction strength similar upon replacing Sr with Ba. This similarity leads to similar behaviors of the two compounds contrary to naive expectation. 

\begin{table} 
\caption{
Effective on-site Coulomb ($U$)/exchange ($J$) interactions between two electrons 
on the same Ir site in the $t_{2g}$ model for all the combinations of Ir-5$d$ orbitals (in eV). 
}
\ 
\label{Table1} 
{\scriptsize
\begin{tabular}{ccccccccccc} 
\hline 
\hline \\ [-8pt]  
Sr$_2$IrO$_4$ &      & $U$   &     &  &     &   & $J$    &    \\ [+1pt]
\hline \\ [-8pt] 
 & $xy$ & $yz$ & $zx$  &  &  & $xy$ & $yz$ & $zx$ \\ 
\hline \\ [-8pt] 
$xy$ & 2.35 & 1.78 & 1.78  & & $xy$ & & 0.16 & 0.16 \\ 
$yz$ & 1.78 & 2.21 & 1.74  & & $yz$ & 0.16 & & 0.15 \\ 
$zx$ & 1.78 & 1.74 & 2.21  & & $zx$ & 0.16 & 0.15 & & \\ 
\hline 
\hline \\ [-8pt]  
Ba$_2$IrO$_4$ &      & $U$   &     &  &     &    &  $J$   &    \\ [+1pt]
\hline \\ [-8pt] 
 & $xy$ & $yz$ & $zx$  &  &  & $xy$ & $yz$ & $zx$ \\ 
\hline \\ [-8pt] 
$xy$ & 1.89 & 1.43 & 1.43  & & $xy$ & & 0.14 & 0.14 \\ 
$yz$ & 1.43 & 1.94 & 1.55  & & $yz$ & 0.14 & & 0.14 \\ 
$zx$ & 1.43 & 1.55 & 1.94  & & $zx$ & 0.14 & 0.14 & & \\ 
\hline 
\hline 
\end{tabular} 
} 
\end{table} 

\paragraph{-{\it DMFT analysis.}}
In DMFT, the inter-atomic correlations are neglected, and the system is mapped self-consistently on an effective impurity model, which we solve with a Monte Carlo method\cite{ctqmc} keeping only the density-density terms of the on-site interaction. We have performed two types of calculations. First, we have investigated the Mott insulator scenario. For this purpose the system was constrained to the paramagnetic phase and the interaction was treated as a free parameter of simplified form $U_{ij} = U$ (this approximation will be justified a posteriori). Second, we studied the real materials with the interaction strengths at their respective first-principles cRPA values (including the orbital dependence). We allowed symmetry breaking towards a long-range magnetic order and studied its evolution as a function of temperature. For all the interaction strengths studied, we have found that the both materials behave effectively as a one-band systems with the $j=1/2$ being half-filled, while the $j=3/2$ bands completely filled. Therefore the $j=1/2$ intra-band interaction is the only parameter, which controls the charge gap opening. 

\begin{figure}[htbp]
\begin{center}
\includegraphics[width=0.49\textwidth]{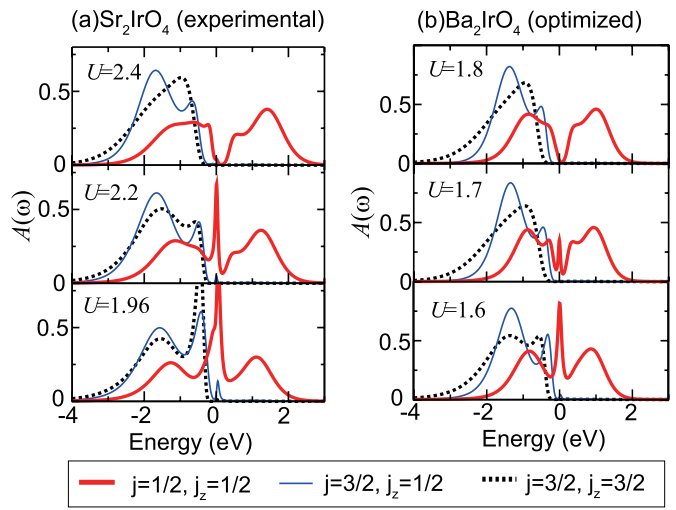}
\caption{(Color online) 
Non-magnetic LDA+DMFT result for Sr$_2$IrO$_4$ (a) and Ba$_2$IrO$_4$ (b) at 
$T=1/40$ eV. 
}
\label{Fig2}
\end{center}
\end{figure} 

One-particle spectra in the paramagnetic phase obtained for various interaction strengths are shown in Fig.~\ref{Fig2}. The metal-insulator transition occurs at $U \sim 2.3 (1.8)$ eV for Sr$_2$IrO$_4$ (Ba$_2$IrO$_4$) while the intra-$j$=1/2 repulsion from cRPA amounts to 1.96 (1.65) eV for Sr$_2$IrO$_4$ (Ba$_2$IrO$_4$). Thus for the cRPA interaction, both compounds remain metallic if paramagnetic. These metals, however, have large mass renormalization factors $Z \sim 3.5 (6)$ for Sr$_2$IrO$_4$ (Ba$_2$IrO$_4$) at $U = U_{\rm cRPA}$ and $T = 1/80$ eV. Here, a rough lower bound of $Z$ is given by $Z = 1 - {\rm Im}\Sigma(i\omega_1)/\omega_1$ at the lowest Matsubara frequency $\omega_1$.

\begin{figure}[htbp]
\begin{center}
\includegraphics[width=0.49\textwidth]{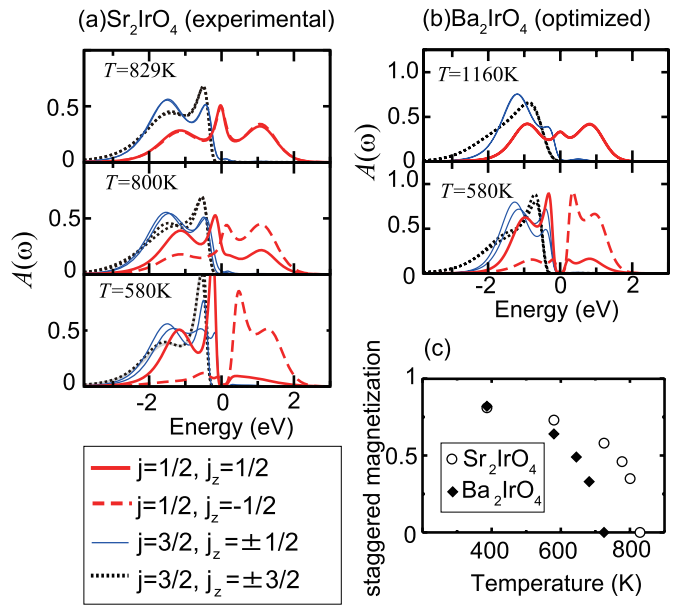}
\caption{(Color online) 
AF LDA+DMFT result for Sr$_2$IrO$_4$ (a) and Ba$_2$IrO$_4$ (b). The Hubbard $U$ estimated by cRPA (1.96 for (a) and 1.6 for (b), respectively) is used. The overestimate of $0.05<T_N<0.1$ in eV in comparison to the experimental values $\sim 250$K may be due to the neglect of the spatial fluctuations in DMFT. Staggered magnetization as a function of temperature is plotted in (c). 
}
\label{Fig3}
\end{center}
\end{figure}

Next, we fix the interaction to the cRPA values. We drop the paramagnetic constraint and allow the system to stabilize an AF order, without altering the Bravais lattice to keep the calculation feasible. Such an in-plane checkerboard order is consistent with the experiment, but the interlayer stacking configuration differs\cite{Takagi}. In Fig.~\ref{Fig3}, we show the result for the one-particle spectra. At high temperatures we obtained metallic paramagnetic solutions, while both of Sr and Ba compounds undergo a transition to the AF phase at around 810 (690) K for the Sr$_2$IrO$_4$ (Ba$_2$IrO$_4$). As the staggered magnetization grows below $T_N$, a gap continuously opens in the one-particle spectra. The continuous transition with a nearby quantum critical point and the resultant enhanced AF fluctuations near it naturally cause the renormalized large-$Z$ metals described above. This is in contrast with the first-order nature claimed in the recent variational Monte-Carlo calculation\cite{Yunoki}. While the calculations predict the two compounds to order at similar temperatures, the actual values are overestimated by a factor 2-3. This is not surprising given the lack of the inter-site fluctuations in DMFT and the layered nature of the materials.

The gap opening can be understood considering the quasiparticle equation 
$
\epsilon_{\sigma}^*-\epsilon_0-\Sigma'_{\sigma}(\epsilon_{\sigma}^*)=0,
$
where $\epsilon_{\sigma}^*$ is the quasiparticle energy, $\epsilon_0$ is the bare energy, $\sigma=\pm$ is the pseudospin index
and $\Sigma'$ is the real part of the self-energy. 
$\Sigma'$ consists of a frequency independent Hartree term and
frequency dependent 'wiggle' at low frequency, responsible for the mass renormalization in PM phase. 
In the traditional Slater picture only the Hartree part of $\Sigma'$ is considered and the gap opens when the $\Sigma'_{+}-\Sigma'_{-}$ splitting
exceeds the bare bandwidth. Here, the Hartree splitting, about 30\% smaller
than the bare bandwidth, is not enough to open the gap. 
The necessary ingredient is the low-frequency 'wiggle' in $\Sigma'_{\sigma}(\omega)$
and its $\sigma$-dependent shift towards positive (negative) frequencies for the empty (filled) orbitals.
Similar $\sigma$-dependent shift in the case of an AF Mott insulator was observed in \cite{sangio}. 
While the Hartree splitting is a simple consequence of a static polarization and does not depend on particular ordering, 
this $\sigma$-dependent shift of $\Sigma'_{\sigma}(\omega)$ is characteristic of 
the AF order, where the empty and filled orbitals on the neighboring sites have the same $\sigma$,
generating a hybridization repulsion between the empty and filled bands. 
The large mass renormalization $Z\ge 3.5-6$ 
present already in the PM phase cooperatively helps the gap opening. In this regard, 
we note that $Z$ is larger than $Z\sim 2$ for SrVO$_3$ and CaVO$_3$, typical 3$d$ correlated metals\cite{Sekiyama}. While DMFT is not expected to predict $T_N$ accurately, it correctly captures the formation of the one-particle gap induced by the developed AF correlations, which in real systems may exist even in some temperature interval above $T_N$. On the other hand, in the absence of SOI, the system is metallic in the LDA+$U$ calculation\cite{KimPRL}. Given the fact that LDA+$U$ usually overestimates the insulating gap, we should indeed have a metallic solution in LDA+DMFT, if we neglect SOI.

\paragraph{-{\it Conclusion.}}
The electronic structures of Sr$_2$IrO$_4$ and Ba$_2$IrO$_4$ calculated from first principles combined with the DMFT show that they have AF and insulating ground states.
These iridates undergo continuous transitions to paramagnetic metals above $T_N$,
indicating that they are to be classified as Slater insulators, in agreement with the available experiments. 
Though they have an essential difference from the Mott insulator, they are not simple Slater insulators either, because strongly renormalized bad metals 
emerge 
in the paramagnetic phase. The present Slater insulators are the consequence of substantial cooperation of Mott-type correlation effects. The strongly-renormalized paramagnetic metal adjacent to the AF Slater insulator opens a possibility of unexplored correlation effects under the interplay of the spin-orbit interaction.
The similarity and dissimilarity to the cuprates elucidated here offers intriguing reference systems to understand the superconducting mechanism in the cuprates when carriers are doped.      

\paragraph{-{\it Acknowledgments}}
We are indebted to J. Akimitsu, H. Okabe, S. Biermann and C. Martins for fruitful discussions, and P. Werner for providing his Monte-Carlo code. We thank financial support from the Grant Agency of the Czech Republic (JK, Grant No. P204/10/0284), NSF (AGE, Grant No. OCI-0904972), FIRST and JST-PRESTO (RA), MEXT Japan (RA and MI, Grant No. 22104010) and Strategic Programs for Innovative Research (SPIRE), MEXT, and the Computational Materials Science Initiative (CMSI), Japan (RA and MI). AVK acknowledges the computational resources of the CSCS and of the NCCS and the CNMS at ORNL, which are sponsored by the respective facilities divisions of the offices of ASCR and BES of the U.S. DoE.


\end{document}